\documentclass[11pt, letter]{article}
\usepackage{fullpage}

\usepackage{belov_paper}

\title{Adaptive Lower Bound for Testing Monotonicity on the Line}
\author{
  Aleksandrs Belovs\thanks{Faculty of Computing, University of Latvia. Raina bulvaris 19, Riga, Latvia, aleksandrs.belovs@lu.lv}
}
\date{}

\begin{document}
\maketitle
\begin{abstract}
In the property testing model, the task is to distinguish objects possessing some property from the objects that are far from it.
One of such properties is monotonicity, when the objects are functions from one poset to another.  This is an active area of research.
In this paper we study query complexity of $\eps$-testing monotonicity of a function $f\colon [n]\to[r]$.  All our lower bounds are for adaptive two-sided testers.
\begin{itemize}
\item We prove a nearly tight lower bound for this problem in terms of $r$.  The bound is $\Omega\sA[\frac{\log r}{\log \log r}]$ when $\eps = 1/2$.
No previous satisfactory lower bound in terms of $r$ was known.
\item We completely characterise query complexity of this problem in terms of $n$ for smaller values of $\eps$.  The complexity is $\Theta\sA[\eps^{-1} \log (\eps n)]$.  
Apart from giving the lower bound, this improves on the best known upper bound.
\end{itemize}

Finally, we give an alternative proof of the $\Omega(\eps^{-1}d\log n - \eps^{-1}\log\eps^{-1})$ lower bound for testing monotonicity on the hypergrid $[n]^d$ due to Chakrabarty and Seshadhri (RANDOM'13).
\end{abstract}

\section{Introduction}

The framework of property testing was formulated by Rubinfeld and Sudan~\cite{rubinfeld:propertyTesting} and Goldreich \etal~\cite{goldreich:propertyTesting}.
A property testing problem is specified by a \emph{property} $\cP$, which is a class of functions mapping some finite set $D$ into some finite set $R$, and \emph{proximity parameter} $\eps$, which is a real number between 0 and 1.
An \emph{$\eps$-tester} is a bounded-error randomised query algorithm which, given oracle access to a function $f\colon D\to R$, distinguishes between the case when $f$ belongs to $\cP$ and the case when $f$ is $\eps$-far from $\cP$.
The latter means that any function $g\in\cP$ differs from $f$ on at least $\eps$ fraction of the points in the domain $D$.
The usual complexity measure is the number of queries to the function $f$.
A tester is with \emph{1-sided} error if it always accepts a function $f$ in $\cP$.
A tester is \emph{non-adaptive} if its queries do not depend on the responses received to the previous queries.
The most general tester is adaptive with 2-sided error, which we will implicitly assume in this paper.

When both the domain $D$ and the range $R$ are partially ordered sets, a natural property to consider is that of \emph{monotonicity}.
A function $f\colon D\to R$ is called \emph{monotone} if $x\le y$ implies $f(x)\le f(y)$ for all $x,y\in D$.
Usually it is assumed that $R$ is a totally ordered set.
In this case, the property $\cP$ consists of all monotone functions from $D$ to $R$. 
The problem of testing monotonicity was explicitly formulated and studied by Goldreich \etal~\cite{goldreich:monotonicityTesting} in the case when $D$ is the Boolean hypercube $\{0,1\}^d$ and $R=\bool$.
This is an active area of research as exemplified by the papers~\cite{dodis:improvedTestingMonotonicity, fischer:monotonicitytesting, chakrabarty:optimalLipschitzHypergrids, chakrabarty:sublinearMonotonicity, chen:newAlgorithmsLowerBoundsMonotonicity, chen:booleanMonotonicityRequiresSqrtn, khot:monotonicityTesting, belovs:monotonicityQuantum, belovs:monotonicityBooleanCube, chen:beyondTalagrand}.

However, slightly earlier than Goldreich \etal, Erg\"un \etal studied the same problem for functions $f\colon [n]\to[r]$.  Erg\"un \etal called it \emph{spot-checker for sorting}, but now this problem is generally known as \emph{monotonicity testing on the line}, the line being the totally ordered set $[n]$.
This problem is the main focus of this paper.
Although this problem is arguably simpler than testing monotonicity on the hypercube, it seems more natural and important from the practical point of view.
Erg\"un \etal mention that their algorithm can be used in software quality assurance by providing a very fast verification procedure that checks whether a presumably sorted array is indeed sorted.

A related problem is that of testing monotonicity on the \emph{hypergrid}.
A hypergrid is a set $[n]^d$ of $d$-tuples with elements in $[n]$.  For two $d$-tuples $x = (x_1,\dots,x_d)$ and $y = (y_1,\dots,y_d)$, we have $x\le y$ iff $x_i\le y_i$ for all $i$.  We are interested in functions from $[n]^d$ to $[r]$ for some positive integers $n,d$ and $r$.
Clearly, this is a generalisation of both monotonicity testing on the line (when $d=1$) and on the hypercube (when $n=2$).
These problems are closely related, so it is important to consider all of them when discussing prior work.

\paragraph{Prior work.}
We proceed with a brief discussion of previous results related to our paper.
Let us start with the upper bounds.
As mentioned above, the problem of testing monotonicity on the line was first considered by Erg\"un \etal~\cite{ergun:spotCheckers}, who gave an $\eps$-tester with complexity $O\sA[\eps^{-1}\log n]$.
Concerning the case of functions from the hypercube $\{0,1\}^d$ to arbitrary $[r]$, Goldreich \etal~\cite{goldreich:monotonicityTesting} proposed the \emph{edge tester}, and Chakrabarty and Seshadhri~\cite{chakrabarty:optimalLipschitzHypergrids} proved that it has query complexity $O(d/\eps)$.
In the latter paper, an $\eps$-tester for testing monotonicity on the hypergrid $[n]^d$ with complexity $O(\eps^{-1}d\log n)$ was also constructed.

Now let us turn to the lower bounds.
Erg\"un \etal~\cite{ergun:spotCheckers} proved a lower bound%
\footnote{If a lower bound does not state dependence on $\eps$, it is assumed that the lower bound holds for some choice of $\eps=\Omega(1)$.}
of $\Omega(\log n)$ for testing monotonicity on the line in the so-called \emph{comparison-based} model.
In this model, each query of the tester may depend only on the order relations between the responses to the previous queries, but not on the values of the responses themselves.\footnote
{
Note that this does \emph{not} mean that the tester asks queries of the form $f(x)\stackrel{?}{\le}f(y)$.  The query is still an input $x$, and the tester learns about the order relations between $f(x)$ and $f(y)$ for \emph{all} previously queried $y$'s.
}
Fischer~\cite{fischer:strengthOfComparisons} proved that any lower bound for a monotonicity testing problem in the comparison-based model implies the same lower bound in the usual value-based model.
This immediately gives an $\Omega(\log n)$ lower bound for testing monotonicity on the line, matching the upper bound by Erg\"un \etal~\cite{ergun:spotCheckers}.
Chakrabarty and Seshadhri~\cite{chakrabarty:OptimalLowerBoundMonotonicity} used the same technique to prove a lower bound of $\Omega(\eps^{-1}d\log n - \eps^{-1}\log\eps^{-1})$ for testing monotonicity on hypergrids.
Unfortunately, Fischer's construction is based on Ramsey theory, which means that, in order for this construction to work, the size of the range, $r$, has to be really huge.

Large values of $r$ in the above lower bounds may lead one to study complexity of testing monotonicity with respect to the size of the range, $r$, rather than the size of the domain, $n$.
Moreover, there are two parameters related to the output of the function $f$: the size of the range (codomain) $r$, and the size of the image $t = \absA|f([n])|$: the number of different values attained by the function.
Clearly, $t\le r$.
This means it is more interesting to prove upper bounds in terms of $t$ and lower bounds in terms of $r$.

Let us start with the upper bounds.
First, it is easy to see that if $r=2$, then $O(1/\eps)$ queries suffice to $\eps$-test monotonicity on the line.
Generalising this observation, Pallavoor \etal~\cite{pallavoor:testingParameterized} constructed an $\eps$-tester for monotonicity on the line with complexity $O(\frac1\eps\log t)$ for general $t$, as well as an $\eps$-tester for monotonicity on hypergrids with complexity $O\sA[\frac d\eps \log\frac d\eps \log t]$.
Since $t\le n$, the lower bound of $\Omega(\log n)$ due to Fischer~\cite{fischer:strengthOfComparisons} implies the lower bound of $\Omega(\log t)$ for all $n\ge t$ and $r$ large enough.%
\footnote{
The size of the domain of a function $f\colon [n]\to[r]$ can be inflated without changing its distance to monotonicity by replacing $f$ with the function $f'\colon [nk]\to[r]$ given by $f'(x) = f(\floor[x/k])$.
}

The main prior technique capable of proving strong lower bounds for functions with small range is that of communication complexity.
Blais \etal introduced this technique in~\cite{blais:testingLowerViaCommunication}, where it was proven that $\Omega\sA[\min\{d, r^2\}]$ queries are required to test a function $f\colon\{0,1\}^d\to[r]$ for monotonicity.
In a subsequent paper~\cite{blais:testingOverHypergrid} a \emph{non-adaptive} lower bound of $\Omega(d\log n)$ was proven for functions $f\colon [n]^d\to[nd]$ on the hypergrid.
In the special case of $d=1$, this gives a lower bound of $\Omega\sA[\log\min\{n,r\}]$ for testing monotonicity of a function $f\colon [n]\to[r]$ on the line.

\paragraph{Our results.}
Our main contribution is an adaptive lower bound for testing monotonicity on the line.  In order not to obstruct our main argument, we first prove the bound for $\eps = \Omega(1)$, and then show how to adapt the construction for smaller values of $\eps$. 
\begin{thm}
\label{thm:main}
Every adaptive bounded-error $1/2$-tester for monotonicity of a function $f\colon [2^k]\to [k^{3k}]$ has query complexity $\Omega(k)$.
\end{thm}

Unlike~\cite{fischer:strengthOfComparisons, chakrabarty:OptimalLowerBoundMonotonicity}, we bypass Ramsey theory and construct two explicit distributions of functions that are hard to distinguish by an adaptive algorithm.
In terms of the size of the domain, $n$, this gives the same lower bound of $\Omega(\log n)$ as in Fischer's paper~\cite{fischer:strengthOfComparisons}, but with vastly reduced range size.
A more direct construction can be beneficial for generalisation to other models, like quantum testers, since it is not known how to adapt Fischer's technique to the quantum settings.

In terms of the size of the range, $r$, this gives a lower bound of of $\Omega\sA[\frac{\log r}{\log \log r}]$, thus nearly matching the upper bound by Pallavoor \etal~\cite{pallavoor:testingParameterized}.
Finally, we get a slightly worse estimate (in terms of the range size) than that of Blais \etal~\cite{blais:testingOverHypergrid} but for \emph{adaptive} testers.
We prove \rf{thm:main} in \rf{sec:main}.

In \rf{sec:smallEpsilon}, we consider the case of general $\eps$.
First, we show that the construction of \rf{thm:main} can be used to prove an $\Omega\sA[\eps^{-1} \log (\eps n)]$ lower bound for $\eps$-testing monotonicity on the line.
For large values of $\eps$, this matches the upper bound of $O\sA[\eps^{-1} \log n]$ due to Erg\"un \etal~\cite{ergun:spotCheckers}, but is slightly worse when $\eps$ is close to $1/n$.
However, we manage to improve the algorithm and prove an upper bound of $O\sA[\eps^{-1} \log (\eps n)]$ for all $\eps n\ge 2$, thus matching our lower bound.
If $\eps n < 2$, the complexity is obviously $\Theta(n)$, hence, this completely resolves the problem of $\eps$-testing monotonicity on the line for all values of $n$ and $\eps\le 1/2$.

Finally, in \rf{sec:hypergrid}, we show how our construction can be adapted to prove the lower bound $\Omega(\eps^{-1}d\log n - \eps^{-1}\log\eps^{-1})$ for $\eps$-testing monotonicity on the hypergrid $[n]^d$.
This coincides with the bound by Chakrabarty and Seshadhri~\cite{chakrabarty:OptimalLowerBoundMonotonicity}, but, again, our construction bypasses Ramsey theory, which results in a vastly reduced range size.

\section{Preliminaries}
Although this is not standard, it will be convenient for us to denote $[n]=\{0,1,\dots,n-1\}$, and $[a..b]=\{a,a+1,\dots,b-1\}$.
Thus, $[n] = [0..n]$, and note that $[a..b]$ does not contain $b$.

An \emph{assignment} $\alpha\colon S\to [r]$ is a function defined on a subset $S\subseteq[n]$.  
The \emph{weight} of $\alpha$ is the size of $S$.
We say that $f$ \emph{agrees} with $\alpha$ if $f(x)=\alpha(x)$ for all $x\in S$.
This is notated by $f\rightharpoondown \alpha$.
The choice of notation is to distinguish it from $f\sim\mu$ which means that $f$ is distributed according to the probability distribution $\mu$.

All logarithms are to the base of 2.

\section{Proof of \rf{thm:main}}
\label{sec:main}

We will define a probability distribution $\mu$ on monotone functions $f\colon [2^k]\to [k^{3k}]$ and a probability distribution $\nu$ on functions $g\colon [2^k]\to [k^{3k}]$ that are $1/2$-far from monotone.
It will be impossible to distinguish these two distributions using fewer than $\Omega(k)$ queries.
Let $m=k^3$, so that $r=m^k$.

We define the distribution $\mu$ in two different but equivalent ways.
First, $\mu$ is defined as the last member in an inductively-defined family $\mu_0,\mu_1, \dots,\mu_k$ of distributions, where $\mu_i$ is supported on functions $[2^i]\to[m^i]$.
The distribution $\mu_0$ is supported on the only function that maps 0 to 0.
Assume that $\mu_i$ is already defined and let us define $\mu_{i+1}$.
In order to do that, we independently sample $f_0$ and $f_1$ from $\mu_i$ and $a$ from $[m-1]$.  The corresponding function $f$ in $\mu_{i+1}$ is given by
\begin{equation}
\label{eqn:f}
f(x) = 
\begin{cases}
a\cdot m^i + f_0(x), & \text{if $0\le x< 2^i$;}\\
(a+1)m^i + f_1(x-2^i), & \text{if $2^i\le x< 2^{i+1}$.}
\end{cases}
\end{equation}

For an alternative way of defining $\mu$, let us assume that the argument $x$ is written in binary and the value $f(x)$ in $m$-ary.
We prepend leading zeroes if necessary so that each number has exactly $k$ digits.
We enumerate the digits from left to right with the elements of $[k]$, so that the $0$-th digit is the most significant one, and the $(k-1)$-st digit is the least significant one.
For each binary string $s$ of length strictly less than $k$, sample an element $a_s$ from $[m-1]$ independently and uniformly at random.
The $i$-th digit of $f(x)$ is defined as $a_s+b$, where $s$ is the prefix of $x$ of length $i$ and $b$ is the $i$-th bit of $x$.
It is easy to see that both definitions of $\mu$ are equivalent, and that any function $f$ from the support of $\mu$ is monotone.

The distribution $\nu$ is defined as the uniform mixture of the following distributions $\nu^j$ for $j\in[k]$.
The function $g\sim\nu^j$ is defined as $g(x) = f(x\oplus 2^{k-1-j})$ when $f$ is sampled from $\mu$.  Here $\oplus$ denotes the bit-wise XOR function.
In other words, the $j$th bit of the argument is flipped before applying $f$.
Alternatively, we may say that $g\sim\nu^j$ is defined as in the case of $\mu$ with the exception that the $j$-th digit of $g(x)$ is $a_s+(1-b)$ instead of $a_s+b$.

\begin{clm}
\label{clm:non-monotone}
Any function $g$ in the support of $\nu^j$ is $1/2$-far from monotone.
\end{clm}

\pfstart
Consider two input strings $x<y$ that differ only in the $j$-th bit.
By the definition of $\nu^j$ we have $g(x)>g(y)$, thus, $\{x,y\}$ is a monotonicity-violating pair.  We have $2^{k-1}$ such disjoint pairs, and every monotone function differs from $g$ on at least one element of each pair.
\pfend

Now we are ready to start with the proof of \rf{thm:main}.  Assume towards contradiction that there exists a randomised $1/2$-tester $\cA$ with query complexity $o(k)$.
Using standard error reduction, we may assume that $\cA$ errs with probability at most $1/8$ on each input.  Let $\lambda$ be the uniform mixture of $\mu$ and $\nu$.
Clearly, $\cA$ errs with probability at most $1/8$ on $\lambda$.
The randomised algorithm $\cA$ can be defined as a probability distribution on deterministic query algorithms of the same query complexity, hence, one of the deterministic algorithms in the support of $\cA$ errs with probability at most $1/8$ on $\lambda$.
Thus, we may assume $\cA$ is a \emph{deterministic} query algorithm.
The error probability $1/8$ on $\lambda$ implies that on $\mu$ and $\nu$ we have
\begin{equation}
\label{eqn:accprob}
\Pr_{f\sim\mu} [\text{$\cA$ accepts $f$}] \ge \frac34
\qquad\text{and}\qquad
\Pr_{g\sim\nu} [\text{$\cA$ accepts $g$}] \le \frac14.
\end{equation}
Also, we may assume that $k$ is large enough, so that the query complexity of the deterministic query algorithm $\cA$ is less than $k/2$.

So, $\cA$ is a decision tree.  Its leaves are specified by assignments in the sense that $\cA$ terminates its work on input $f$ in a leaf given by assignment $\alpha$ if and only if $f$ agrees with $\alpha$.  Moreover, the weight of each such $\alpha$ does not exceed the query complexity of $\cA$.
We partition all these assignments $\alpha$ into two parts as follows.
We say that an integer $\ell\in [r]$ is \emph{good} if it contains no digit 0 and no digit $m-1$ (in the $m$-ary representation as before).  Otherwise, $\ell$ is \emph{bad}.
We say that an assignment $\alpha$ is \emph{good} if all the elements in its image are good.  Otherwise, $\alpha$ is \emph{bad}.
Finally, a leaf of $\cA$ is good iff the corresponding assignment is good.

\begin{clm}
\label{clm:bad}
The probability that $\cA$ ends its work in a bad leaf when run on an input $f$ sampled from $\mu$ is $o(1)$.
\end{clm}

\pfstart
For each $x$ in the domain of $f$, the value $f(x)$ is bad only if one of the $k$ elements $a_{s_0},a_{s_1},\dots,a_{s_{k-1}}$ has value 0 or $m-2$, where $s_i$ is the prefix of $x$ of length $i$.
Thus, the probability of $\cA$ to terminate in a bad leaf is at most the probability of finding an element $a_s$ with value in $\{0,m-2\}$ with $k/2$ queries, where on each query it is allowed to test the values of $k$ different $a_s$'s.
Since the $a_s$'s are independent, and the probability of $a_s\in\{0,m-2\}$ is $2/(m-1)$, we get, using the standard bound on search, that the probability of succeeding is at most
$\frac k2\cdot k\cdot \frac{2}{m-1} = o(1)$.
\pfend

\mycommand{fa}{\Pr_{f\sim\mu} [f\rightharpoondown \alpha]}
\mycommand{ga}{\Pr_{g\sim\nu} [g\rightharpoondown \alpha]}

\begin{clm}
\label{clm:good}
For each good assignment $\alpha$ of weight at most $k/2$,
\[
\fa\le 2\cdot\ga .
\]
\end{clm}

Before we start with the proof of this claim, let us show how \rf{thm:main} follows from Claims~\ref{clm:bad} and~\ref{clm:good}.
In the following, let $C$ be the set of assignments which correspond to the accepting leaves of $\cA$, let $B\subseteq C$ be the subset of bad assignments, and $G\subseteq C$ be the subset of good assignments.
Then,
\begin{align*}
\Pr_{f\sim\mu} [\text{$\cA$ accepts $f$}] 
= \sum_{\alpha\in B}\fa + \sum_{\alpha\in G}\fa&\\
\le o(1) + 2\sum_{\alpha\in G}\ga &\le o(1) + 2\Pr_{g\sim\nu} [\text{$\cA$ accepts $g$}],
\end{align*}
which is in contradiction with~\rf{eqn:accprob} if $k$ is large enough.
\medskip

It remains to prove \rf{clm:good}.
Consider a good assignment $\alpha$ of weight at most $k/2$.  Let $S$ be the domain of $\alpha$.
We say that a pair $x<y$ from $S$ \emph{cuts} an index $j\in[k]$ iff their first $j$ bits agree, and they disagree in the $j$-th bit.  In other words, there exists $a\in\bZ$ such that 
\[
2a \cdot2^{k-j-1} \le x< (2a+1)\cdot2^{k-j-1}\le y <(2a+2)\cdot2^{k-j-1}.
\]
The assignment $\alpha$ cuts all the indices cut by the pairs in $S$.
\rf{clm:good} follows from the following two lemmata.

\begin{lem}
\label{lem:goodalpha}
If a good assignment $\alpha$ does not cut an index $j$, then
\[
\Pr_{f\sim\mu} [f\rightharpoondown\alpha]
=
\Pr_{g\sim\nu^j} [g\rightharpoondown\alpha].
\]
\end{lem}

\pfstart
By induction on $i$ in the definition~\rf{eqn:f} of $\mu_i$.
Let for brevity $j' = k-j-1$.
If $j'<i$, we define $\nu^j_i$ as the distribution over the functions $g(x) = f(x\oplus 2^{j'})$ when $f$ is sampled from $\mu_i$.  If $j'\ge i$, we define $\nu^j_i = \mu_i$.
We prove that 
\begin{equation}
\label{eqn:agree}
\Pr_{f\sim\mu_i} [f\rightharpoondown\alpha]
=
\Pr_{g\sim\nu^j_i} [g\rightharpoondown\alpha]
\end{equation}
for every good assignment $\alpha$ from $[2^i]$ to $[m^i]$ that does not cut the index $j$.

The base case $i=0$ is trivial.  (Actually, the statement is trivial for all $i\le j'$.)
Assume~\rf{eqn:agree} is proven for $i$, and let us prove it for $i+1$.
Let $S$ be the domain of $\alpha$.  There are two cases.

First, assume both $S\cap[2^i]$ and $S\cap [2^i..2^{i+1}]$ are non-empty.
This means that $\alpha$ cuts $k-i-1$, hence, $i\ne j'$.
Also, we may assume there exists $1\le a\le m-3$ such that  $\alpha([2^i])\subseteq [am^i.. (a+1)m^i]$ and $\alpha([2^i..2^{i+1}])\subseteq [(a+1)m^i..(a+2)m^i]$, since otherwise both sides of~\rf{eqn:agree} are 0.
Under these assumptions,
\[
\Pr_{f\sim\mu_{i+1}} [f\rightharpoondown\alpha]
=
\frac1{m-1}
\Pr_{f_0\sim\mu_i} [f_0\rightharpoondown\alpha_0]
\Pr_{f_1\sim\mu_i} [f_1\rightharpoondown\alpha_1],
\]
where $f_0$ and $f_1$ are obtained reversely from~\rf{eqn:f}, and $\alpha_0$ and $\alpha_1$ are defined similarly:
$\alpha_0(i) = \alpha(i)\bmod m^i$ and $\alpha_1(i) = \alpha(i+2^i)\bmod m^i$ for all $i\in [2^i]$.
Both assignments are good, and they do not cut the index $j$.
Similarly, since $i\ne j'$:
\[
\Pr_{g\sim\nu^j_{i+1}} [g\rightharpoondown\alpha]
=
\frac1{m-1}
\Pr_{g_0\sim\nu^j_i} [g_0\rightharpoondown\alpha_0]
\Pr_{g_1\sim\nu^j_i} [g_1\rightharpoondown\alpha_1].
\]
By the inductive assumption, we have the required equality.

Now assume one of $S\cap[2^i]$ and $S\cap [2^i..2^{i+1}]$ is empty.
We consider the case $S\subseteq[2^i]$, the second one being similar.
Again, we can assume there exists $1\le a\le m-2$ such that  $\alpha([2^i])\subseteq [am^i.. (a+1)m^i]$.  Then,
\[
\Pr_{f\sim\mu_{i+1}} [f\rightharpoondown\alpha]
=
\frac1{m-1}
\Pr_{f_0\sim\mu_i} [f_0\rightharpoondown\alpha_0]
\]
and, no matter whether $i=j'$ or not,
\[
\Pr_{g\sim\nu^j_{i+1}} [g\rightharpoondown\alpha]
=
\frac1{m-1}
\Pr_{g_0\sim\nu^j_i} [g_0\rightharpoondown\alpha_0],
\]
and again we have the required equality by the inductive assumption.
\pfend

\begin{lem}
\label{lem:cut}
An assignment $\alpha$ of weight $t$ cuts at most $t-1$ indices in $[k]$.
\end{lem}

\pfstart
Let $S$ be the domain of $\alpha$, and let $J\subseteq[k]$ be the set of indices that $\alpha$ cuts.

Let us construct a graph $G$ as follows.  Its vertex set is $S$.  For every index $j\in J$ take one arbitrary pair of elements $x,y\in S$ that cuts $j$ and connect $x$ and $y$ by an edge.  We say that the edge $xy$ cuts $j$.

We claim that the graph $G$ is acyclic, from which the statement of the lemma follows.
Assume that $G$ contains a simple cycle.  Consider an edge $xy$ that cuts the minimal index $j$ on this cycle.  Then $x$ and $y$ disagree in the $j$-th bit.  On the other hand, considering the remaining part of the cycle, we see that $x$ and $y$ agree in the $j$-th bit.  A contradiction, hence, $G$ is acyclic.
\pfend

By \rf{lem:cut}, there are at least $k/2$ indices not cut by $\alpha$, and using \rf{lem:goodalpha}, we have
\[
2\cdot\ga \ge \frac 2k \sum_{j:\text{$\alpha$ does not cut $j$}} \Pr_{g\sim \nu^j} [g\rightharpoondown\alpha] \ge \frac 2k\cdot \frac k2\cdot \fa = \fa,
\]
proving \rf{clm:good}.

\section{The Case of Small $\eps$}
\label{sec:smallEpsilon}

In this section, we briefly describe how the result of \rf{thm:main} can be extended to arbitrary values of $\eps$, and give an improved version of the algorithm by Erg\"un \etal~\cite{ergun:spotCheckers}.

\begin{thm}
\label{thm:witheps}
If $\eps\le 1/2$, the complexity of $\eps$-testing a function $f\colon[n]\to [r]$ for monotonicity is 
\[
\Omega\s[\min \sfigB{\frac{\log(\eps n)}{\eps},\, \frac{\log (\eps r)}{\eps\log\log (\eps r)}}].
\]
\end{thm}

\mycommand{tmu}{\widetilde{\mu}}
\mycommand{tnu}{\widetilde{\nu}}

\pfstart
The proof closely follows that of \rf{thm:main}.  We will briefly describe the construction and the proof using the notation of \rf{sec:main}.

Let $\ell$ be a positive integer and assume $\eps=1/(2\ell)$.  We will construct probability distributions $\tmu$ and $\tnu$ on functions $f\colon [\ell 2^k]\to [\ell k^{3k}]$ such that all functions in the support of $\tmu$ are monotone,  functions in the support of $\tnu$ are $\eps$-far from monotone, and it takes $\Omega(\ell k)$ queries to distinguish $\tmu$ and $\tnu$.
Expressing $\ell k$ in terms of $\eps$ and $n$ or in terms of $\eps$ and $r$ gives the required bound.

Let $\mu$ and $\nu$ be as in \rf{sec:main}.  For $s\in[\ell]$, independently sample $f_s$ from $\mu$.  Define $f\sim\tmu$ as
\begin{equation}
\label{eqn:tmu}
f(s\cdot 2^k+x) = s\cdot k^{3k} + f_s(x)
\end{equation}
for all $s\in[\ell]$ and $x\in[2^k]$.
The distribution $\tnu$ is defined as the uniform mixture of $\tnu^{t,j}$ as $t$ ranges over $[\ell]$ and $j$ over $[k]$.  The corresponding function $f\sim \tnu^{t,j}$ is defined as in~\rf{eqn:tmu} with exception that $f_t$ is sampled from $\nu^j$ instead of $\mu$.
It is easy to see that functions in the support of $\tmu$ are monotone, and, using \rf{clm:non-monotone}, that functions in the support of $\tnu$ are $\eps$-far from monotone.

Informally, it takes $\Omega(\ell k)$ queries to distinguish $\tmu$ and $\tnu$ because we are searching for one non-monotone distribution $\nu^j$ among $\ell$ independent distributions.
Formally, we may proceed as follows.  Assume towards contradiction that there exists a deterministic query algorithm $\cA$ that makes less than $\ell k/4$ queries, accepts $\tmu$ with probability at least $3/4$ and accepts $\tnu$ with probability at most $1/4$.

Using the same reasoning as in \rf{clm:bad}, the expected number of bad elements found by $\cA$ when run on $\tmu$ is $O(\ell k^2/m) = o(\ell)$.  We call an assignment $\alpha$ on $[\ell 2^k]$ \emph{bad} if it has more than $\ell/4$ bad elements in its image.  Otherwise, we call $\alpha$ good.  By Markov's inequality, the probability $\cA$ terminates in a bad assignment when executed on $\tmu$ is $o(1)$.

Now consider a good assignment $\alpha$ of weight at most $\ell k/4$.
It corresponds to $\ell$ sub-assignments $\alpha_s$ on $[2^k]$ defined by $\alpha_s(x) = \alpha(s\cdot2^k+x) \bmod k^{3k}$.
Using \rf{clm:good}, we get that
\begin{equation}
\label{eqn:notcutting}
\Pr_{f\sim \tmu} [f\rightharpoondown\alpha] = \Pr_{g\sim \tnu^{t,j}} [g\rightharpoondown\alpha]
\end{equation}
if the sub-assignment $\alpha_t$ is good and does not cut $j$.  Using that there are at most $\ell/4$ bad $\alpha_s$ and \rf{lem:cut}, we have that there are at least $\ell k/2$ pairs $(t,j)$ satisfying~\rf{eqn:notcutting}.
Hence,
\[
\Pr_{f\sim \tmu} [f\rightharpoondown\alpha] \le 2 \cdot \Pr_{g\sim \tnu} [g\rightharpoondown\alpha].
\]
Now we finish the proof as in \rf{sec:main}.
\pfend

\begin{thm}
Assume $\eps n\ge 2$.  
Then, there exists a non-adaptive 1-sided algorithm that tests a function $f\colon[n]\to [r]$ for monotonicity using $O\sA[\frac{\log(\eps n)}{\eps}]$ queries.
\end{thm}

Combined with the result of \rf{thm:witheps}, we get that complexity of this problem is $\Theta\sA[\frac{\log(\eps n)}{\eps}]$ if $\eps n\ge 2$, and $\Theta(n)$ otherwise.

\pfstart
The algorithm is inspired by that of Erg\"un \etal~\cite{ergun:spotCheckers}.
\[
\fbox{ \parbox{.9\textwidth}{
\begin{enumerate}
    \item Repeat $\Theta(1/\eps)$ times:
    \negmedskip
    \begin{enumerate}
        \item Choose $x\in[n]$ uniformly at random.
        \item Query $f(x)$.
        \item For $i=0,\dots,\ceil[\log (\eps n)]$:
        \begin{itemize}
            \item Let $w$ be the largest multiple of $2^i$ strictly smaller than $x$, and $y$ be the smallest multiple of $2^i$ strictly larger than $x$.  (If any of them is outside $[n]$, do not use it on the next steps.)
            \item Query $f(w)$ and $f(y)$.
            \item If $f(w)>f(x)$ or $f(x)>f(y)$, reject the function $f$.
        \end{itemize}
    \end{enumerate}
    \negmedskip
    \item If no contradiction to monotonicity was found, accept.
\end{enumerate}
}}
\]

Clearly, the query complexity of the algorithm is $O\sA[\frac{\log(\eps n)}{\eps}]$, it is non-adaptive, and it always accepts a monotone function $f$.
Assume now that $f$ is $\eps$-far from monotone.

\begin{lem}
If $f$ is $\eps$-far from monotone, there exists a collection of pairwise disjoint pairs $(x_1,y_1),\dots,(x_t, y_t)$ for $t = \floor[\eps n/2]$ such that, for all $i$,
$x_i<y_i$, $f(x_i)>f(y_i)$, and $y_i - x_i \le \eps n$.
\end{lem}

\pfstart
The pairs can be constructed using the following algorithmic procedure.
Start with $S\gets [n]$ and $i\gets 1$.  We treat $S$ as a sorted list.
While $i\le t$, choose two neighbouring elements $x_i<y_i$ in $S$ such that $f(x_i)>f(y_i)$, remove $x_i$ and $y_i$ from $S$, and increment $i$.

It remains to prove that (a) such a pair $(x_i,y_i)$ will always exist and (b) that $y_i-x_i\le \eps n$.
For (a), observe that $i\le t$ implies that we have removed strictly less than $\eps n$ elements from $S$ so far.  
Since $f$ is $\eps$-far from monotone, we have that $f$ restricted to $S$ is not monotone (otherwise, it would be possible to extend $f|_S$ to a monotone function on all $[n]$).
The existence of the pair $(x_i,y_i)$ is now obvious.

For (b), again, observe that we have removed less than $\eps n-1 $ elements from $S$ so far.  The elements $x_i$ and $y_i$ are neighbouring in $S$, which means they are at distance at most $\eps n$ in the original list $[n]$.
\pfend

\begin{lem}
Assume $x$ and $y$ satisfy $x<y$, $f(x)>f(y)$ and $y-x\le \eps n$.
Then one element $z\in\{x,y\}$ of these two is such that the algorithm will reject if it chooses $z$ on step 1(a).
\end{lem}

\pfstart
If $y = x+1$, the algorithm will reject if it chooses either of $x$ or $y$.
So, assume $y\ge x+2$.

We claim that there exists an integer $0\le i\le \ceil[\log(\eps n)]$ such that there is unique multiple of $2^i$ strictly between $x$ and $y$.  
Indeed, there is at least one for $i=0$ and at most one for $i = \ceil[\log(\eps n)]$.  Also, it is not possible that there is more than one for some value of $i$ and zero for $i+1$.

Let $w$ be this unique multiple of $2^i$.  
If $f(w) < f(x)$, then it will be detected if the algorithm chooses $x$ on step 1(a).
If $f(w) \ge f(x)$, then $f(w) > f(y)$, and it will be detected if the algorithm chooses $y$ on step 1(a).
\pfend

By the previous two lemmata, there exist $\floor[\eps n/2]$ values of $x$ on which the algorithm rejects should it choose any of them on step 1(a).
The probability of choosing one of them on one iteration of the loop is $\Omega(\eps)$.  The probability to choose any of them on one of the $\Theta(1/\eps)$ iterations of the loop is $\Omega(1)$.
\pfend

\section{The Case of Hypergrids}
\label{sec:hypergrid}
In this section, we show how our results can be transformed into a lower bound for testing monotonicity on the hypergrid.

\begin{thm}
If $\eps\le 1/2$, the complexity of $\eps$-testing a function $f\colon[n]^d\to [r]$ for monotonicity is 
\[
\Omega\s[\min \sfigB{\frac{\log(\eps n^d)}{\eps},\, \frac{\log (\eps r)}{\eps\log\log (\eps r)}}].
\]
\end{thm}

In terms of $n$, the lower bound can be expressed as $\Omega(\eps^{-1}d\log n - \eps^{-1}\log \eps^{-1})$.

\pfstart
Consider the functions defined in the proof of \rf{thm:witheps}.
Assume that $\ell = 2^a$ for some integer $a$, and choose $k$ so that $a+k = db$ for some integer $b$.
We consider the domain of the input functions $[\ell 2^k] = [2^{a+k}]$ as $[2^b]^d$, where we break the binary representation of $x\in[2^{a+k}]$ into $d$ groups of $b$ bits.

If a function is monotone on $[2^{a+k}]$, it is still monotone when considered on $[2^b]^d$.  
Also, the analysis of the algorithm does not involve the order relation defined on the domain of the input functions.  
The only thing that might go wrong is that the functions in the support of $\tnu$ are no longer $\eps$-far from monotone when considered on $[2^b]^d$.  But this does not happen.  
Indeed, in the proof of \rf{clm:non-monotone}, the monotonicity-violating pairs $(x,y)$ differ in exactly one bit.  So if $x<y$ in $[2^{a+k}]$, this is still true in $[2^b]^d$, hence the proof of \rf{clm:non-monotone} carries over.
Thus, all the functions in the support of $\tnu$ are $\eps$-far from monotone.
The statement of the theorem now follows from the proof of \rf{thm:witheps}.
\pfend

\section*{Acknowledgements}
I am grateful to Eric Blais for introducing me to this problem, and for his encouragement to work on it, as well as for pointing out Ref.~\cite{pallavoor:testingParameterized}.
I would like to thank Dmitry Gavinsky, Ansis Rosmanis, and Ronald de Wolf for helpful discussions, and anonymous referees for their suggestions.
The construction of \rf{thm:witheps} is due to Ronald de Wolf.

This research is supported by the ERDF project number 1.1.1.2/I/16/113.
Part of this work was done while visiting Institute of Mathematics of the Czech Academy of Sciences in Prague, and Centre for Quantum Technologies in Singapore.  I would like to thank Dmitry Gavinsky, Pavel Pudl\'ak, and Miklos Santha for hospitality.

\small
\bibliographystyle{habbrvM}
\bibliography{belov}

\end{document}